\documentclass[showpacs,twocolumn,nofootinbib]{revtex4-1}
\usepackage[english]{babel}
\usepackage{graphicx}
\usepackage{amsmath, amssymb}
\usepackage{dcolumn}
\usepackage{subfigure}

\newcommand{\be}{\begin{equation}}
\newcommand{\ee}{\end{equation}}
\newcommand{\bq}{\begin{eqnarray}}
\newcommand{\eq}{\end{eqnarray}}
\newcommand{\Dslash}{\hbox{$\partial\!\!\!{\slash}$}}

\newcommand{\bslash}{\hbox{$b\!\!\!{\slash}$}}

\newcommand{\kbruto}{\hbox{$k \!\!\!{\slash}$}}
\newcommand{\pbruto}{\hbox{$p \!\!\!{\slash}$}}

\newcommand{\bbruto}{\hbox{$b \!\!\!{\slash}$}}

\begin{document}

\title{\bf Arbitrariness in the gravitational Chern-Simons-like term induced radiatively}

\author{J. C. C. Felipe $^{a, d}$}\email[]{guaxu@fisica.ufmg.br}
\author{A. R. Vieira $^a$}\email[]{arvieira@fisica.ufmg.br}
\author{A. L. Cherchiglia $^a$}\email[]{adriano@fisica.ufmg.br}
\author{A. P. Ba\^eta Scarpelli $^b$}\email[]{scarpelli.apbs@dpf.gov.br}
\author{Marcos Sampaio $^{a, c}$}\email[]{msampaio@fisica.ufmg.br}

\affiliation{$^a$ Universidade Federal de Minas Gerais - Departamento de F\'{\i}sica - ICEX \\ P.O. BOX 702,
30.161-970, Belo Horizonte MG - Brazil}
\affiliation{$^b$ Setor T\'{e}cnico-Cient\'{\i}fico - Departamento de Pol\'\i cia Federal,\\
Rua Hugo D'Antola, 95 - Lapa - S\~{a}o Paulo - Brazil}
\affiliation{$^c$ Centre for Particle Theory - Dept of Mathematical Sciences - Durham University\\
Science Laboratories South Rd. Durham DH1 3LE - UK}
\affiliation{$^d$ School of Mathematics and Statistics - Newcastle University\\
Newcastle upon Tyne NE1 7RU - UK}


\begin{abstract}

The induction of a Lorentz- and CPT-violating Chern-Simons-like term in a fermionic theory embedded in linearized quantum gravity is reassessed.
We explicitly show that gauge symmetry on underlying Feynman diagrams  does not fix the arbitrariness inherent to such induced term at one loop order. We present the calculation in a nonperturbative expansion in the Lorentz-violating parameter $b_\mu$ and within a framework which,  besides operating in the physical dimension,  judiciously parametrizes regularization dependent arbitrary parameters usually fixed by symmetries.

\end{abstract}

\maketitle

\section{Introduction}
\label{s1}

In the Standard Model of particle physics, Lorentz and CPT are regarded as fundamental symmetries. However, since the early 90's possible violations of such symmetries have been studied \cite{Kostelecky}-\cite{Lehner2}. The first model, introduced by Sean M. Carroll et al \cite{Carroll}, considered the theoretical and phenomenological consequences of adding to QED a Chern-Simons-like term proportional to a constant four-vector. They found out that such model predicts vacuum birefringence. However, astrophysical data establish stringent bounds to this kind of deviations from Lorentz and CPT symmetries \cite{{Carroll},{Goldhaber}}. Such small effects would come from spontaneous symmetry breaking
of Lorentz symmetry in a more complete theory such as string theory \cite{Kostelecky}.

One interesting aspect which has been vastly investigated is whether this CS-like term can be radiatively produced. One example of this mechanism occurs in extended QED with a Lorentz violating axial term, in which the CS-like term appears when we consider radiative corrections to the photon propagator. However, different results for the CS-like coefficient have been found (see, for example, \cite{{Altschul},{Colladay},{Perez},{Perez2},{Jackiw},{Chung},{Bonneau}, {Klinkhamer1},{Jackiw3},{Scarpelli}}). The coefficient of the induced term, coming from the cancelations of divergences, is in fact regularization dependent. Using Pauli-Villars method, for instance, this coefficient is found to be zero \cite{{Colladay},{Jackiw3}}, while the result using dimensional regularization depends on  how the dimensional continuation of the $\gamma_5$ matrix is carried out  (see \cite{{Tsai},{Tsai2}} and references therein).

Following the idea of an induced CS-like term in extended QED, it has also been discussed if a gravitational CS-like term can be
radiatively induced in a fermionic theory in curved space-time. Phenomenologically, the existence of such term would imply that gravitational waves possess two degrees of polarization instead of four\cite{Jackiw2}. Nevertheless, the coefficient of such induced term turns out to depend on details involving the regularization of intermediate divergences as well \cite{{Mariz}, {Mariz2}}.

In this work, we compute the 1-loop correction to the graviton propagator in the weak field approximation, using a more general approach called Implicit Regularization. Since it does not specify any particular regularization technique, allowing the reproduction of other results by choosing the method at the end of the calculation, it permits us to identify the sources of ambiguities. We find that the induced gravitational CS-like term depends on a set of surface terms which, coming from differences of divergent integrals, are arbitrary.

Following \cite{Jackiw3}, arbitrary parameters that appear in finite radiative corrections must be fixed either by phenomenology or symmetries of the underlying model. By demanding gauge invariance of the action, which enforces transversality of the graviton self-energy, we find that the dependence in one of the surface terms remains in the final amplitude.  This is the same result obtained in the case of extended QED in flat space. In such case, requiring transversality of the final amplitude does not determine the coefficient of the Carroll-Field-Jackiw term.

The paper is organized as follows: in section \ref{s2}, we carry out, with a pedagogical purpose, a review of the calculation of the induced CS-like term in extended QED in flat space. In section \ref{s3}, we turn our attention to fermions in linearized quantum gravity with a Lorentz violating extension. We compute the 1-loop correction to the graviton propagator with Implicit Regularization to study the induced CS-like term in this case. In section \ref{s4}, we conclude and leave details of the integrals that appear in this work to appendix \ref{A}.

\section{Revisiting the induction of a CS-like term in extended QED}
\label{s2}

In order to motivate our line of reasoning, we revisit the induction of the Chern-Simons-like
term (also called Carroll-Field-Jackiw term) in extended QED, whose action reads
\be
S_{QED}^{ext} = \int d^4x \, \bar{\psi} (i \partial\!\!\!/ - A\!\!\!/ - m - b\!\!\!/ \gamma_5
) \psi.
\ee
The coefficient of the induced CFJ term is well known to be ambiguous and many different methods have been applied, furnishing various results. Here, we take as an example the calculation of \cite{Scarpelli}, in which the Implicit Regularization scheme has been used. For simplicity, we treat the massless case. If the fermion is  non-massive, its propagator can be decomposed as \cite{Altschul}
\be
\frac{i}{\kbruto -\bbruto \gamma_5} = \frac{i}{\kbruto-\bbruto}P_L+ \frac{i}{\kbruto+\bbruto}P_R,
\label{proj}
\ee
where we are using the chiral projectors
\be
P_{R,L}=\frac {1\pm \gamma_5}{2}.
\ee
Note that with this decomposition, it is simple to perform the complete one loop calculation, without necessity of expanding the propagator. So, it is really a nonperturbative calculation in $b_\mu$ and the problem reduces to the calculation of just one Feynman graph. Here, we  carry out the calculation with an arbitrary loop routing. The full one-loop photon self-energy is given by
\be
\Pi^{\mu \nu}=\frac 12 \left\{ \Pi^{\mu \nu}_+ + \Pi^{\mu \nu}_- + \Pi^{\mu \nu}_{5+}+\Pi^{\mu \nu}_{5-}\right\},
\ee
with
\bq
&&\Pi^{\mu \nu}_{\pm}(p,\alpha p \pm b) = \nonumber \\
&&\int_k^\Lambda \mbox{tr}\left\{ \frac{\gamma^\nu(\kbruto + \alpha \pbruto \pm \bbruto) \gamma^\mu
\left[\kbruto+ (\alpha+1)\pbruto \pm \bbruto\right]}{(k +\alpha p \pm b)^2\left[k+(\alpha+1)p \pm b\right]^2} \right\}
\eq
and
\bq
&&\Pi^{\mu \nu}_{5\pm}(p,\alpha p \pm b) = \nonumber \\
&&\pm \int_k^\Lambda \mbox{tr}\left\{ \frac{\gamma^\nu(\kbruto + \alpha \pbruto \pm \bbruto) \gamma^\mu
\left[\kbruto+ (\alpha+1)\pbruto \pm \bbruto\right]\gamma_5}{(k +\alpha p \pm b)^2\left[k+(\alpha+1)p \pm b\right]^2} \right\},
\eq
where $\int_k \equiv \int \frac{d^4k}{(2 \pi)^4}$ and the superscript $\Lambda$ is used to indicate that some four dimensional regularization
has been applied (say a cutoff) just to justify algebraic operations at the level of the integrands. Since the regularization was not specified yet, we can maintain, for a while, the dependence on the parameter $\alpha$. For a particular momentum routing in the loop, a variable $\alpha$ is fixed. This is just illustrative, since the dependence on $\alpha$ cannot be disentangled from the choice of the regularization procedure.

The induction of the CS-term comes from the $\Pi^{\mu \nu}_{5\pm}$ parts, so that we have
\bq
\Pi^{\mu \nu}_{5} && =\frac 12 \left[ \Pi^{\mu \nu}_{5+}(p,\alpha p +b)+\Pi^{\mu \nu}_{5-}(p,\alpha p -b)\right] \nonumber \\
&& = \frac 12\left[\Pi^{\mu \nu}_{5+}(p,b_1)-\Pi^{\mu \nu}_{5+}(p,b_2)\right],
\eq
with $b_1=\alpha p+b$ and $b_2=\alpha p-b$. So, let us calculate $\Pi^{\mu \nu}_{5+}(p,b)$,
which, after Dirac algebra, can be written as
\bq
\Pi^{\mu \nu}_{5+}(p,b)&&=4ip_\beta \epsilon^{\nu \alpha \mu \beta}
\int_k^\Lambda \frac{(b+k)_\alpha}{(k+b)^2(k+p+b)^2} \nonumber \\
&&=4ip_\beta \epsilon^{\nu \alpha \mu \beta}(b_\alpha I + I_\alpha),
\label{CFT}
\eq
with
\be
I,I_\alpha=\int_k^\Lambda \frac {1,k_\alpha}{(k+b)^2(k+p+b)^2}.
\label{int}
\ee

We apply the Implicit Regularization framework \cite{IReg} to treat these integrals. Let us make a brief review of the method. In this scheme, we assume the existence of an implicit regulator ($\Lambda$) in order to judiciously use the following identity to separate  UV divergent basic integrals from the finite part:
\begin{align}
\int_k\frac{1}{(k+p)^2-m^2}&=\int_k\frac{1}{k^2-m^2}\nonumber\\
&-\int_k\frac{(p^2+2p\cdot k)}{(k^2-m^2)[(k+p)^2-m^2]},
\label{9}
\end{align}
where we have introduced a fictitious mass in the propagators. This is necessary because, although  the present integrals are infrared safe, the above expression without mass will break the original integral in two infrared divergent parts. The limit $m^2 \to 0$ is taken in the end. In this process a renormalization scale $\lambda \ne 0$ is introduced. In general, besides a finite part in the UV limit, we get basic divergent integrals which are defined as
\be
I^{\mu_1 \cdots \mu_{2n}}_{log}(m^2)\equiv \int_k \frac{k^{\mu_1}\cdots k^{\mu_{2n}}}{(k^2-m^2)^{2+n}}
\ee
and
\be
I^{\mu_1 \cdots \mu_{2n}}_{quad}(m^2)\equiv \int_k \frac{k^{\mu_1}\cdots k^{\mu_{2n}}}{(k^2-m^2)^{1+n}}.
\ee

The basic divergences with Lorentz indices can be judiciously combined as differences between integrals with the same superficial degree of divergence, according to the equations below, which define surface terms\footnote{The Lorentz indices between brackets stand for symmetrization of the tensor, i.e. $A^{\{\alpha_1\cdots\alpha_n}B^{\beta_1\cdots\beta_n\}}=A^{\alpha_1\cdots\alpha_{n}}B^{\beta_1\cdots\beta_n}$ + sum over permutations between the two sets of indices $\alpha_1\cdots\alpha_{n}$ and $\beta_1\cdots\beta_n$.}:
\begin{align}
\Upsilon^{\mu \nu}_{2w}=& \eta^{\mu \nu}I_{2w}(m^2)-2(2-w)I^{\mu \nu}_{2w}(m^2) \nonumber \\
\equiv& v_{2w}\eta^{\mu \nu},
\label{dif1}\\
\nonumber\\
\Xi^{\mu \nu \alpha \beta}_{2w}=& \eta^{\{ \mu \nu} \eta^{ \alpha \beta \}}I_{2w}(m^2)\nonumber \\
&-4(3-w)(2-w)I^{\mu \nu \alpha \beta }_{2w}(m^2) \nonumber \\
\equiv& \xi_{2w}\eta^{\{ \mu \nu} \eta^{ \alpha \beta \}},
\label{dif2}\\
\nonumber\\
\Sigma^{\mu \nu \alpha \beta \gamma \delta}_{2w} =& \eta^{\{\mu \nu} \eta^{ \alpha \beta} \eta^ {\gamma \delta \}}I_{2w}(m^2)\nonumber\\
&-8(4-w)(3-w)(2-w) I^{\mu \nu \alpha \beta \gamma \delta}_{2w}(m^2) \nonumber \\
\equiv& \sigma_{2w} \eta^{\{\mu \nu} \eta^{ \alpha \beta} \eta^ {\gamma \delta \}}.
\label{dif3}
\end{align}

In the expressions above, $2w$ is the degree of divergence of the integrals and  for the sake of brevity, we substitute the subscripts $log$ and $quad$ by $0$ and $2$, respectively. Surface terms can be conveniently written as integrals of total derivatives, namely
\bq
&&v_{2w}\eta^{\mu \nu}= \int_k\frac{\partial}{\partial k_{\nu}}\frac{k^{\mu}}{(k^2-m^2)^{2-w}}, \nonumber \\
\label{ts1}
\eq
\begin{align}
(\xi_{2w}-v_{2w})\eta^{\{ \mu \nu} \eta^{ \alpha \beta \}}= \int_k\frac{\partial}{\partial k_{\nu}}\frac{2(2-w)k^{\mu} k^{\alpha} k^{\beta}}{(k^2-m^2)^{3-w}},
\label{ts2}
\end{align}
and
\bq
&&(\sigma_{2w}-\xi_{2w})\eta^{\{ \mu \nu} \eta^{ \alpha \beta} \eta^ {\gamma \delta \}} =\nonumber \\
&&\int_k\frac{\partial}{\partial k_{\nu}}\frac{4(3-w)(2-w)k^{\mu} k^{\alpha} k^{\beta} k^{\gamma} k^{\delta}}{(k^2-m^2)^{4-w}}.
\label{ts3}
\eq

We see that equations (\ref{dif1})-(\ref{dif3}) are undetermined because they are differences between divergent quantities. Each regularization scheme gives a different value for these terms. However, as physics should not depend on the schemes applied, we leave these terms to be arbitrary until the end of the calculation, fixing them by symmetry constraints or phenomenology, when it applies.

Concerning the surface terms, a comment is in order. As is well known, to perform shifts in integrals with degree of divergence which are at least linear, it is necessary to compensate with surface terms. For this reason, in a 4D procedure as Implicit Regularization, which preserves until the end the surface terms, the final amplitude will depend on the routing in the loop momentum. This dependence appears in the coefficients of the surface terms. Nevertheless, in Implicit Regularization scheme, the parameters defined in equations (\ref{dif1})-(\ref{dif3}) are adjusted in order to fix symmetries.

Returning to our calculations, the results of the integrals (\ref{int}) in the Implicit Regularization framework are given by
\be
I=I_{log}(\lambda^2)-\frac{i}{16 \pi^2}\left[ \ln{ \left(-\frac{p^2}{\lambda^2}\right)}-2\right]
\ee
and
\bq
I_\alpha =&&-\frac{(p+2b)_\alpha}{2} \left\{ I_{log}(\lambda^2) \right. \nonumber \\
&& \left. -\frac{i}{16 \pi^2}\left[ \ln{ \left(-\frac{p^2}{\lambda^2}\right)}-2\right] -v_0\right\} \nonumber \\
=&&-\frac{(p+2b)_\alpha}{2}\left(I-v_0\right).
\eq
Substituting these results in equation (\ref{CFT}), we get
\be
\Pi^{\mu \nu}_{5+}(p,b)=4 i v_0 b_\alpha p_\beta \epsilon^{\nu \alpha \mu \beta}.
\ee
So, we obtain
\bq
\Pi^{\mu \nu}_{5} && =\frac 12 \left[4i v_0 (\alpha p+b)_\alpha p_\beta \epsilon^{\nu \alpha \mu \beta} -
4i v_0 (\alpha p-b)_\alpha p_\beta \epsilon^{\nu \alpha \mu \beta} \right] \nonumber \\
&& = 4i v_0 b_\alpha p_\beta \epsilon^{\nu \alpha \mu \beta}.
\eq
The induced coefficient of the Carroll-Field-Jackiw term will then be given by
\be
\Delta c_\mu =2i v_0 b_\mu.
\ee

We see that the coefficient of the induced CS-type term is proportional to the undetermined parameter $v_0$.
In \cite{Jackiw} it was obtained a definite result for $\Delta c_\mu$ in the nonperturbative approach. For this, a procedure was used in the calculation of the surface terms. Actually, these terms are dependent on the procedure adopted. In our result, this is expressed in the dependence on $v_0$.

In \cite{Altschul}, it was shown that the procedure of \cite{Jackiw} has as a consequence the violation of gauge symmetry at second order in $b_\mu$. However, in the follow-up paper \cite{Altschul2}, the author has shown that the use of an adequate Pauli-Villars regulator in the calculation preserves gauge symmetry in second order in $b_\mu$ even in the nonperturbative approach. Enforcing this result, in \cite{Scarpelli}, the complete one-loop calculation was performed with Implicit Regularization. The results for the zeroth and second order terms in $b_\mu$ are given below:
\bq
&&\Pi_{0}^{\mu \nu}=\Pi(p^2)(p^\mu p^\nu-p^2 \eta^{\mu \nu}) -4 v_2 \eta^{\mu \nu}+\nonumber \\
&&-\frac 43 \left\{ v_0(p^\mu p^\nu-p^2 \eta^{\mu \nu})+ \right. \nonumber \\
&& \left. +(2p^\mu p^\nu +p^2 \eta^{\mu \nu})(\xi_0-2 v_0)\right\}
\eq
and
\bq
&&\frac 12\left(\Pi_{bb-}^{\mu \nu}+\Pi_{bb+}^{\mu \nu}\right) \nonumber \\
&& = -4\left\{ \left(b^2\eta^{\mu \nu} +2 b^\mu b^\nu \right)(\xi_0-2v_0) \right\}.
\eq
If one uses symmetric integration when calculating $v_0$ and $\xi_0$, such that $k^\mu k^\nu \to \eta^{\mu \nu}k^2/4$
and $k^\mu k^\nu k^\alpha k^\beta \to \eta^{\{\mu \nu}\eta^{\alpha \beta\}}k^4/24$, one obtains
\be
v_0=\frac {i}{32\pi^2} \;\;\;\; \mbox{and} \;\;\;\; \xi_0= \frac {5i}{96\pi^2},
\ee
so that
\be
\frac 12\left(\Pi_{bb-}^{\mu \nu}+\Pi_{bb+}^{\mu \nu}\right)=
\frac{i}{24 \pi^2}\left( b^2\eta^{\mu \nu} +2 b^\mu b^\nu \right),
\ee
as in \cite{Altschul}. However, this will cause gauge symmetry violation even in the zeroth order term. The condition for transversality of the
photon self-energy for all orders in $b_\mu$ is $\xi_0=2 v_0$ and $v_2=0$. A gauge invariant procedure will respect these conditions, as, for example, the Pauli-Villars regulator used in \cite{Altschul2}. Since the $v_0$ parameter cannot be fixed, the coefficient of the Chern-Simons-like term is really regularization dependent.

\section{Arbitrariness in the induced CS gravity term}
\label{s3}

We consider a massless fermionic theory in a gravitational background with a CPT-violating term,
\begin{equation}
S=\int d^4 x \left(\frac{i}{2}e\ e^{\mu}_{a} \bar{\psi} \gamma^{a} \overleftrightarrow{D}_{\mu} \psi-e\ e^{\mu}_{a}{\it b}_{\mu} \bar{\psi} \gamma^{a} \gamma_5 \psi\right),
\label{1}
\end{equation}
where $e^{\mu}_{a}$ is the tetrad, $e=det\ e^{\mu}_{a}$ and $b_{\mu}$ is a constant four-vector.

In equation (\ref{1}), in order to couple fermions with the gravitational field, we need to define the covariant derivative,
\begin{equation}
D_{\mu} \psi= \partial_{\mu} \psi +\frac{1}{2}\omega_{\mu a b}\sigma^{ab} \psi,
\label{2}
\end{equation}
where $\omega_{\mu a b}$ is the spin connection, which depends on the tetrad, and $\sigma^{ab}=\frac{1}{4}[\gamma^a,\gamma^b]$.

\begin{widetext}

\begin{figure}[!h]
 \includegraphics[scale=0.9]{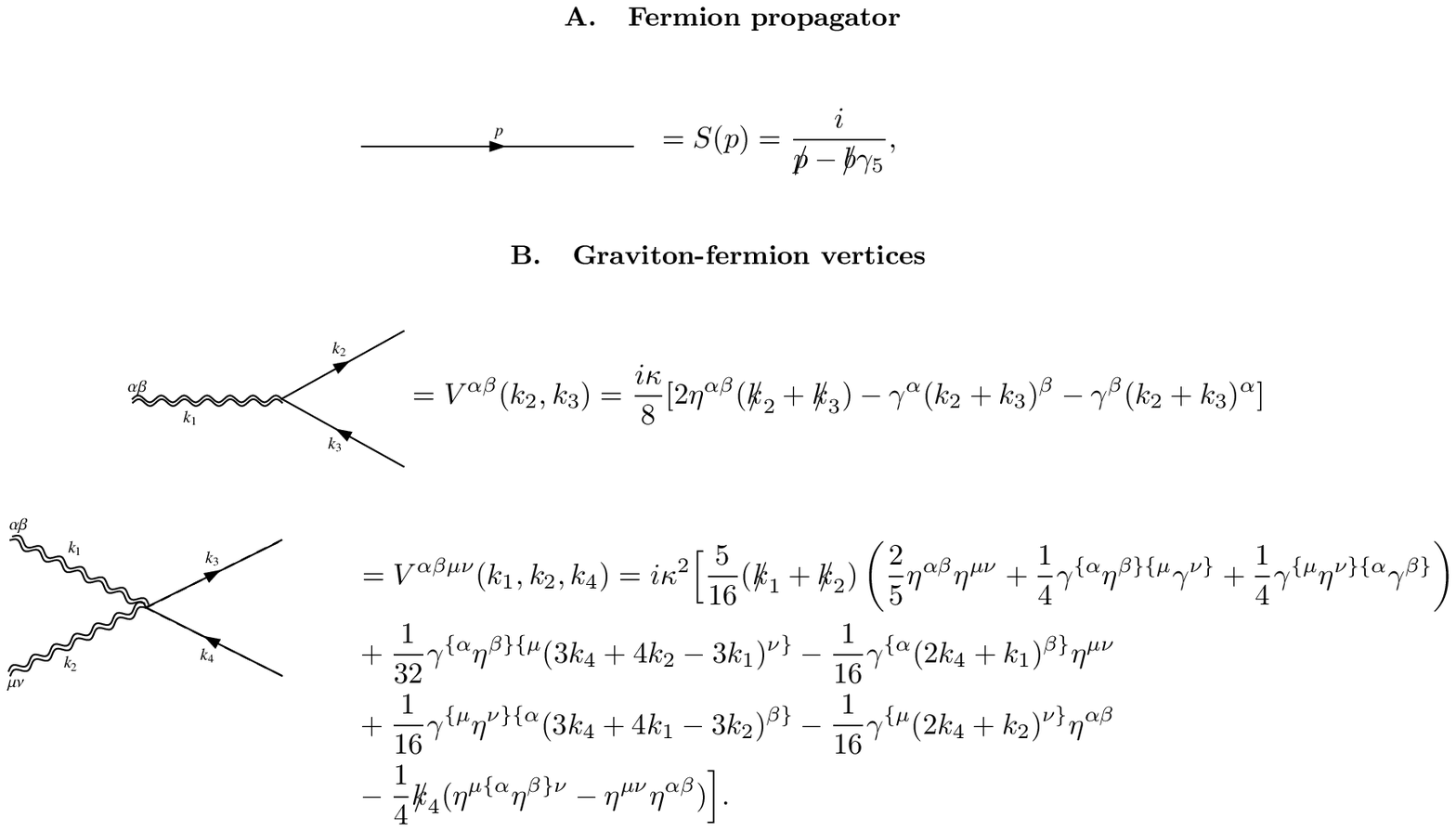}
 \caption{Feynman rules for a fermionic theory in linearized quantum gravity with a Lorentz violating extension.}
 \label{fey}
\end{figure}

\end{widetext}

In the weak field approximation, we use the following expansions for the metric and the tetrad:
\be
g^{\mu\nu}=\eta^{\mu\nu}+\kappa h^{\mu\nu}
\ee
and
\be
e_{\mu a}= \eta_{\mu a}+\frac{1}{2}{\kappa}h_{\mu a}.
\ee
Therefore, the action (\ref{1}) can be reexpressed as
\begin{align}
S=&\int d^4x \Bigg\{\frac{1}{2}i \bar{\psi}\overleftrightarrow{\Dslash}\psi+\frac{1}{2}i\kappa\Big[h\bar{\psi}\overleftrightarrow{\Dslash}\psi
-\frac{1}{2}h^{\mu}_a\gamma^{a}\bar{\psi}\overleftrightarrow{\Dslash}\psi+\nonumber\\
&+\frac{1}{4}\bar{\psi}\partial_{b}h_{ca}\gamma^{\{a}\gamma^{b}\gamma^{c\}}\psi-\frac{1}{4}\bar{\psi}\partial_{c}h_{ba}\gamma^{\{a}\gamma^{b}\gamma^{c\}}\psi \nonumber\\
&+\bar{\psi}h^{\mu}_a b_{\mu}\gamma^{a} \gamma_5\psi+\frac{1}{2}\bar{\psi}h\bslash\gamma_5\psi\Big]-
\bar{\psi}\bslash\gamma_5 \psi \Bigg\}+O(\kappa^2).
\label{3}
\end{align}
Feynman rules, shown in figure \ref{fey}, can be readily derived from equation (\ref{3}).

\begin{figure}[!h]
\centering
    \subfigure[]
    {\includegraphics[trim= 0mm 20mm 0mm 25mm, clip, scale=0.5]{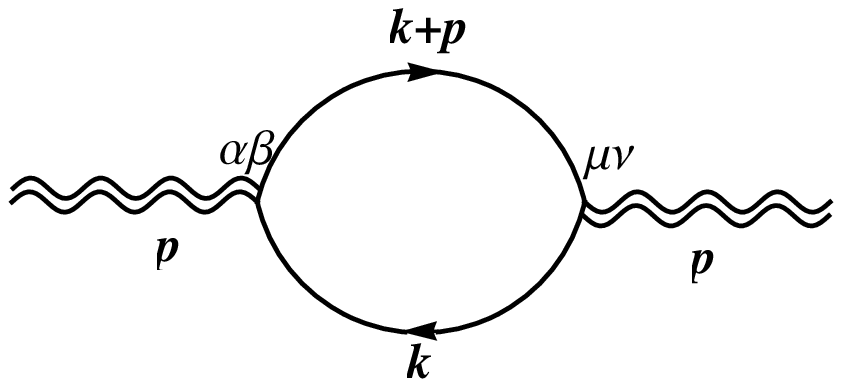}
        \label{figa}}											
		\subfigure[]
{   \includegraphics[trim= 0mm 30mm 0mm 20mm, clip, scale=0.5]{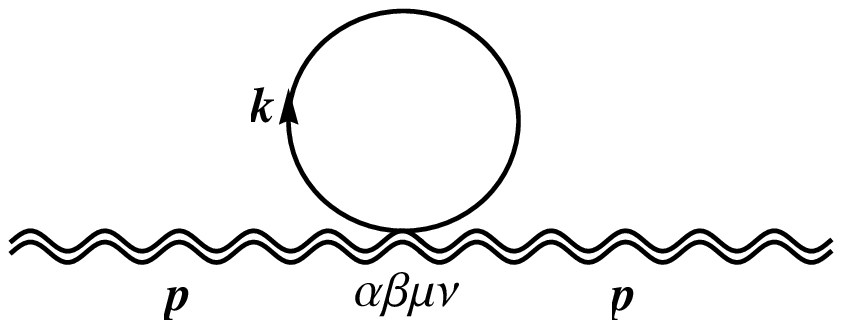}
        \label{figb}}								
				\caption{1-loop corrections for the graviton propagator. The double waved line and the solid line stand for the graviton and the fermion, respectively.}
\label{fig}
\end{figure}

In order to obtain the induced Chern-Simons-like term, we have to compute the linear part in $b_{\mu}$ of the one-loop correction for the graviton propagator. We opt to use the complete propagator rather than treat the axial term as a interaction. Figure \ref{fig} shows the two diagrams that contribute.

\noindent
Their amplitudes read
\begin{align}
\Pi^{\mu \nu \alpha \beta}_{(a)}(p)= i\int \frac{d^4 k}{(2\pi)^4} Tr[&V^{\mu \nu}(k+p,k)S(k+p)\times\nonumber\\
&V^{\alpha \beta}(k,k+p)S(k)]
\end{align}
and
\begin{align}
&\Pi^{\mu \nu \alpha \beta}_{(b)}(p)= i\int \frac{d^4 k}{(2\pi)^4} Tr[S(k)V^{\alpha \beta \mu \nu}(p,p,k)].\nonumber\\
\end{align}

We write the following expansion for the fermion propagator
\bq
\frac{i}{\kbruto -\bbruto \gamma_5} = \sum_{n=0}^\infty \frac{i}{\kbruto}
\left\{ -i \bbruto \gamma_5 \frac{i}{\kbruto} \right\}^n = \sum_{n=0}^\infty S_n(k).
\eq
Since the CS-like term we are interested in is linear in $b_\mu$, we can write
\bq
\Pi^{\mu \nu \alpha \beta}_{(a)CS}(p)= i\int \frac{d^4 k}{(2\pi)^4} Tr[&&V^{\mu \nu}(k+p,k)S_0(k+p)\times\nonumber\\
&&V^{\alpha \beta}(k,k+p)S_0(k)\bslash \gamma_5 S_0(k)] \nonumber \\
+ i\int \frac{d^4 k}{(2\pi)^4} Tr[&&V^{\alpha \beta}(k,k+p)S_0(k)\times\nonumber\\
V^{\mu \nu}(k+p,k)&&S_0(k+p)\bslash \gamma_5 S_0(k+p)]
\label{8.1}
\eq
and
\begin{align}
&\Pi^{\mu \nu \alpha \beta}_{(b)CS}(p)= i\int \frac{d^4 k}{(2\pi)^4} Tr[S_0(k)\bslash \gamma_5 S_0(k)V^{\alpha \beta \mu \nu}(p,p,k)].
\label{8.3}
\end{align}

These amplitudes are symmetric under the exchange $\mu\leftrightarrow \nu$ and $\alpha\leftrightarrow \beta$ as they should. The amplitude $\Pi^{\mu \nu \alpha \beta}_{(b)CS}(p)$ is null after the trace operation. The amplitude $\Pi^{\mu \nu \alpha \beta}_{(a)CS}(p)$ is superficially cubically divergent.

The result of the implicitly regularized amplitude $\Pi^{\mu \nu \alpha \beta}_{(a)CS}(p)$ is given by (see a list of results of integrals in the appendix)

\begin{align}
\Pi^{\mu \nu \alpha \beta}_{(a)CS}(p)= \frac{-i}{8}\kappa^2 \Big[&\left(\frac{i}{48\pi^2}-64\sigma_0-4\upsilon_0+4\xi_0 \right) p^{\alpha}p^{\nu}\nonumber\\
&-\left(\frac{i}{48\pi^2}+32\sigma_0\right) \eta^{\alpha \nu} p^2 \Big]\epsilon^{\lambda \rho \beta \mu}b_{\lambda}p_{\rho}\nonumber\\
+(\alpha \leftrightarrow &\beta)+(\mu \leftrightarrow \nu)+(\alpha \leftrightarrow \beta,\mu \leftrightarrow \nu).
\label{27}
\end{align}	

Obviously this result contains arbitrariness expressed by surface terms. To try to fix them, we demand gauge invariance of the action, expressed by the transversality of the final amplitude. Explicitly, we have
\begin{align}
p_{\alpha}\Pi^{\mu \nu \alpha \beta}_{(a)CS}(p)= \frac{-i}{8}&\kappa^2(4\xi_0-4\upsilon_0-96\sigma_0)\times\nonumber\\
&(\epsilon^{\lambda \rho \beta \mu}p^{\nu}+\epsilon^{\lambda \rho \beta \nu}p^{\mu})p^2 b_{\lambda}p_{\rho}=0
\label{idWard}
\end{align}

In order to satisfy equation (\ref{idWard}), we must have $\xi_0=\upsilon_0= \sigma_0=0$ or $\xi_0-\upsilon_0=24\sigma_0$. The former condition determines the CS-like term and the latter does not. If we replace this expression in equation (\ref{27}) the result is

\begin{align}
\Pi^{\mu \nu \alpha \beta}_{(a)CS}(p)=& \frac{-i}{24}\kappa^2 \epsilon^{\lambda \rho \beta \mu}b_{\lambda}p_{\rho} \left(\!\!\frac{i}{16\pi^2}+96\sigma_0\!\!\right)( p^{\alpha}p^{\nu}-\eta^{\alpha \nu} p^2)\nonumber\\
&+(\alpha \leftrightarrow \beta)+(\mu \leftrightarrow \nu)+ (\alpha \leftrightarrow \beta,\mu \leftrightarrow \nu).
\label{31}
\end{align}

We see that transversality is not sufficient to fix all surface terms leaving us an arbitrary result.  Depending on the choice of the arbitrary term $\sigma_0$, we can either recover other results found in the literature \cite{{Mariz},{Mariz2},{Marcelo}} or even get zero.

The four terms of equation (\ref{31}) assure the symmetry of the amplitude under the change $\mu\leftrightarrow \nu$ and $\alpha\leftrightarrow \beta$. The consequent CS-like effective action is

\begin{align}
\mathcal{L}_{CS}=\!\left(\!\frac{1}{96\pi^2}-16i\sigma_0\!\right)\!\kappa^2
b^{\lambda} h^{\mu \nu}\epsilon_{\alpha \mu \lambda \rho}\partial^{\rho}(\partial^{2}h^{\alpha}_{\nu}-\partial_{\nu} \partial_{\gamma}h^{\gamma \alpha}).
\label{34}
\end{align}

If we set $\sigma_0=0$, this result agrees with the one of reference \cite{Mariz} where dimensional regularization was employed. Such behavior should be expected since the surface terms are zero if explicitly evaluated by this technique.

One more comment is in order. In the case of the extended QED in flat space-time, the transversality of the vacuum polarization tensor is trivially respected by the Carroll-Field Jackiw (CFJ) term, because of the presence of only one antisymmetric L\'evi-Civit\`a tensor contracted with the external momentum. In that case, this symmetry was not an alternative to try to determine the remaining surface term. The case of the Lorentz-violating model in a gravitational background is different, since the satisfaction of this symmetry is not trivial. It was necessary to enforce a relation among three parameters so as to satisfy it.

\section{Concluding remarks}
\label{s4}

In this work, we study the induction of a CS-like term by radiative corrections for a massless Lorentz- and CPT-violating fermionic theory embedded in a curved spacetime. We adopt the framework of Implicit Regularization, which clearly parametrizes regularization dependent terms. Besides, we carry out the calculations in the nonperturbative approach in the Lorentz-violating parameter $b_\mu$. We imposed transversality of the amplitude as an attempt to fix the coefficient of the induced Lorentz-violating term. However, after enforcing this symmetry, the relation to be satisfied by the surface terms are not sufficient to determine the coefficient of the induced CS gravity term, leaving a free parameter.

This result should be compared with the one of the induction of a CFJ term in the extended QED in flat space. In that case, the satisfaction of transversality of the amplitude is trivial due to products involving symmetric and antisymmetric tensors. This is not the case here, since it was necessary to enforce a relation among three parameters so as to satisfy this symmetry.

\begin{widetext}

\section{Appendix} \label{A}

The result of the regularized integrals, after taking the trace, are:

\begin{align}
\int_k \frac{k^2}{k^4(k+p)^2}=& I_{log}(\lambda^2)+2\tilde{b} -\tilde{b} \ln (-\frac{p^2}{\lambda^2}),\\
\int_k \frac{k^2 k^{\alpha}}{k^4(k+p)^2}=& \frac{1}{2}p^{\alpha}\left[-I_{log}(\lambda^2)+\upsilon_0-2\tilde{b} +\tilde{b} \ln \left(-\frac{p^2}{\lambda^2}\right)\right],\\
\int_k \frac{k^{\alpha}k^{\beta}}{k^4(k+p)^2}=& \frac{1}{4}\eta^{\alpha \beta}\left[I_{log}(\lambda^2)-\upsilon_0+2\tilde{b} -\tilde{b} \ln \left(-\frac{p^2}{\lambda^2}\right)\right]+\frac{1}{2}\tilde{b}\frac{p^{\alpha}p^{\beta}}{p^2},\\
\int_k \frac{k^{\mu}k^{\alpha}k^{\beta}}{k^4(k+p)^2}=& \frac{1}{12}p^{\{ \mu}\eta^{\alpha \beta\}}\left[-I_{log}(\lambda^2)+\xi_0+\tilde{b} \ln \left(-\frac{p^2}{\lambda^2}\right)-\frac{5}{3}\tilde{b}\right]-\frac{1}{3}\tilde{b}\frac{p^{\mu}p^{\alpha}p^{\beta}}{p^2},\\
\int_k \frac{k^2 k^{\alpha}k^{\beta}}{k^4(k+p)^2}=& -\frac{1}{4}\eta^{\alpha \beta}p^2[I_{log}(\lambda^2)-\upsilon_0]+\frac{1}{6}(p^2\eta^{\alpha \beta}+2p^{\alpha}p^{\beta})[I_{log}(\lambda^2)-\xi_0]+\nonumber\\
&+\frac{1}{2}\tilde{b} p^2\eta^{\alpha \beta}\left[\frac{1}{6} \ln \left(-\frac{p^2}{\lambda^2}\right)-\frac{4}{9}\right]-\tilde{b} p^{\alpha}p^{\beta}\left[\frac{1}{3}\ln \left(-\frac{p^2}{\lambda^2}\right)-\frac{13}{18}\right],\\
\int_k \frac{k^{\mu}k^{\nu} k^{\alpha}k^{\beta}}{k^4(k+p)^2}=& -\frac{1}{24}\eta^{\{\mu \nu}\eta^{\alpha \beta \}}p^2[I_{log}(\lambda^2)-\xi_0]+\frac{1}{48}(p^2\eta^{\{\alpha \beta}\eta^{\mu \nu \}}+p^{\{\alpha}p^{\beta}\eta^{\mu \nu \}})[I_{log}(\lambda^2)-\xi_0-24\sigma_0]\nonumber\\
&+\frac{1}{8} \tilde{b} p^2\eta^{\{\mu \nu}\eta^{\alpha \beta \}}\left[\frac{1}{6} \ln \left(-\frac{p^2}{\lambda^2}\right)-\frac{4}{9}\right]-\frac{1}{72}\tilde{b} p^{\{\alpha}p^{\beta}\eta^{\mu \nu \}}\left[\frac{3}{2}\ln \left(-\frac{p^2}{\lambda^2}\right)-\frac{5}{2}\right]+\frac{1}{4}\tilde{b}\frac{p^{\alpha}p^{\beta}p^{\mu}p^{\nu}}{p^2},
\end{align}
\begin{align}
\int_k \frac{k^{2}k^{\mu} k^{\alpha}k^{\beta}}{(k^4(k+p)^2}=& \frac{1}{6}p^{\{\mu}\eta^{\alpha \beta \}}p^2[I_{log}(\lambda^2)-\xi_0]-\frac{1}{8}\left(p^2 p^{\{\mu}\eta^{\alpha \beta \}}+2p^{\alpha}p^{\beta}p^{\mu}\right)[I_{log}(\lambda^2)-\xi_0-24\sigma_0]-\nonumber\\
&-\frac{1}{4}\tilde{b} p^2p^{\{\mu}\eta^{\alpha \beta \}}\left[\frac{1}{6} \ln \left(-\frac{p^2}{\lambda^2}\right)-\frac{4}{9}\right]+\tilde{b} p^{\alpha}p^{\beta}p^{\mu}\left[\frac{1}{4}\ln \left(-\frac{p^2}{\lambda^2}\right)-\frac{7}{12}\right],
\end{align}
where $\lambda$ is mass scale and $\tilde{b} \equiv \frac{i}{(4\pi)^2}$.

\end{widetext}
	
{\bf Acknowledgments}

The authors acknowledge fruitful discussions with M. C. Nemes during the preparation of this work. A. L. C. acknowledges financial support by FAPEMIG. M. S. and A. P. B. S. acknowledge research grants from CNPq. J. C. C. F. and A. R. V. acknowledge financial support by CNPq.

\end{document}